\documentclass[showpacs,amsmath,amssymb]{revtex4}

\usepackage{graphicx}
\usepackage{graphics}
\usepackage{dcolumn}
\usepackage{bm}

\usepackage{texdraw}
\usepackage{makeidx,amscd,amsmath,amssymb}
\usepackage{floatflt}
\usepackage{rotating}
\usepackage{array,multirow,color}
\usepackage{url}

\usepackage[latin1]{inputenc}

\begin{document}

\preprint{draft}

\title{Band structure of NiO revisited}

\author{Luiz G. Ferreira}
\email{guima00@gmail.com}
\affiliation{Instituto de F\'{\i}sica, Universidade de S\~{a}o Paulo, 
05315-970 S\~{a}o Paulo, Brazil}
\author{Lara K. Teles}
\email{lkteles@ita.br}
\author{Marcelo Marques}
\email{mmarques@ita.br}
\affiliation{Instituto Tecnol\'{o}gico de Aeron\'{a}utica, 12228-900  S\~{a}o
Jos\'{e} dos Campos, Brazil}

\begin{abstract}
The band structure of a strongly correlated semiconductor as NiO has been
the object of much debate [PRL 103, 036404 (2009); PRL 102, 226401 (2009)].
Most authors, using computational techniques well beyond the simple 
{\it density functional theory} and the approximations GGA or  LDA, claim
that the band gap is about 4.0 eV and that the conduction band is of
Ni3d nature. Thus they seem to forget the results of electron energy-loss
spectroscopy and inelastic x-ray scattering, both able to determine
 electronic transitions of only about 1.0 eV to an optically forbidden
Ni3d band. Further, a 
simple atomic calculation of the Ni$^{++}$ spin flip energy 
demonstrates that a Ni3d band at 4.0 eV is impossible. To set the issue 
straight, we calculated NiO with the very successful technique of
PRB 78, 125116 (2008). It turns out that a band at 4.0 eV is
optically accessible and made of excited atomic states, not Ni3d. Aside
from that, we also found a narrow Ni3d band at about 1.0~eV. To
confirm our procedures once again, we also calculated MnO and obtained
the standard results of the good calculations as those cited above,
and of experiment.

\end{abstract}

\pacs{71.10.-w; 71.15.-m; 71.15.Qe; 71.20.Nr; 71.27.+a}
\maketitle
The band structure of NiO is being the object of many calculations by different
techniques. To quote just a few, Physical Review Letters recently 
published  the
works of Engel and Schmid \cite{Engel} on the Exact-Exchange technique, and of
 Tran and Blaha \cite{Tran} on a novel semilocal 
exchange-correlation potential.
Physical Review published the work of R\"{o}dl et al \cite{Rodl}  on a GGA
\cite{Perdew}
calculation.  The Journal of Physics: Condensed Matter published a
self-interaction (SIC) calculation \cite{Dane}. The many authors claim that
NiO presents an important challenge
 for density functional theory because it is a strongly correlated system. 
The most
simple LSDA and GGA calculations are reported to be 
failures. Thus, along the years
NiO is being calculated with many sophisticate techniques. Aside from
the works already quoted, we can cite some examples:  
Deng et al used a variational LDA formulation \cite{Deng},
Fuchs et al \cite{Fuchs} with their generalized Kohn-Sham method, 
Kobayashi et al 
\cite{Kobayashi} that used a GW started from LSDA+U,  Eder \cite{Eder} 
that used a Variational Cluster Approximation, Miura et al \cite{Miura},
Kune\v{s} et al \cite{Kunes}, and Ren et al \cite{Ren} using 
different LDA+DFMT, Tran et al \cite{Blaha} with a hybrid exchange-correlation,
mixing Hartree-Fock, Han et al \cite{Han} with LDA+U, Li et al \cite{Li}
and Kotani and van Schilfgaarde \cite{Kotani} using a self-consistent GW,
Korotin et al \cite{Korotin} with a Wannier function method.
\par Of course it is impossible to review the many papers. We want only to
call attention that many of these calculations are wrong when they set the 
band gap at about 4.3 eV, following the experimental BIS result of
Sawatzky and Allen \cite{Sawatzky}, claiming that the conduction band
at that energy is a Ni-d band. The location of the conduction Ni-d bands
has been determined by M\"{u}ller and H\"{u}fner \cite{Muller} using
electron energy-loss spectroscopy, and by Huotari et al \cite{Huotari} using
inelastic x-ray scattering. The band at 4.3 eV above valence is certainly not
Ni-d because these start at a much lower energy, about 1.0 eV.  
It is simple to show the approximate position of the Ni-d conduction bands
by calculating the following atomic energy differences  
\[Ni[4s^2(3d\uparrow)^4(3d\downarrow)^4]-
Ni[4s^2(3d\uparrow)^5(3d\downarrow)^3]\]
 or the Ni$^{++}$ ionic difference
\[Ni[(3d\uparrow)^4(3d\downarrow)^4]-Ni[(3d\uparrow)^5(3d\downarrow)^3]\]
 These differences are the energies required to
flip a spin, which correspond to the excitation of an electron 
to the antiferromagnetic crystal conduction band.  Using the 
PBE exchange-correlation \cite{Perdew}, the difference is 0.99~eV
for the atom, and 1.03~eV for the ion. Using the LDA exchange-correlation
of Cepperley-Alder Perdew-Zunger \cite{Zunger} the differences are only 
0.05~eV smaller. In any case, the spin-flip energy is similar to the 
threshold to the first energy band, as found experimentally \cite{Muller,
Huotari} and the 4.3~eV BIS band \cite{Sawatzky} could never be a
Ni-d band. Compared with MnO, also an antiferromagnetic oxide, NiO is very
different. In the case of MnO, the spin-flip energy
\[Mn[4s^2(3d\uparrow)^4(3d\downarrow)^1)]-
Mn[4s^2(3d\uparrow)^5)]\]
is 3.55~eV, similar to the band gap, so that a Mn-d conduction band
is possible. 
\par Next we present the results of our calculation for NiO. Our calculation 
was all-electron with wavefunctions expanded in LAPW+lo 
\cite{WIEN2k}. We avoided the
pseudopotential calculations for two reasons. First, in the case  of
NiO, the Ni-d functions are not easily expanded with plane waves. 
Second, for magnetic systems, the exchange-correlation potential
may depend much on the way the core density is treated \cite{Cohen},
because linearization is unacceptable. So it is not just a problem of 
finding the good pseudopotential, but the good core density must also be
found. NiO is an antiferromagnet with the rock salt structure.
The Ni sublattice is antiferromagnetically ordered as an $L1_1$ (CuPt)
alloy. It has a very small trigonal distortion along the body
diagonal. Along this direction, the planes alternate spins, up and down.
We used the experimental lattice parameters and did not optimize
geometry, which is the best procedure to calculate the elementary
excitation spectrum.
\par
\begin{figure}[htbp]
\includegraphics[width=9cm]{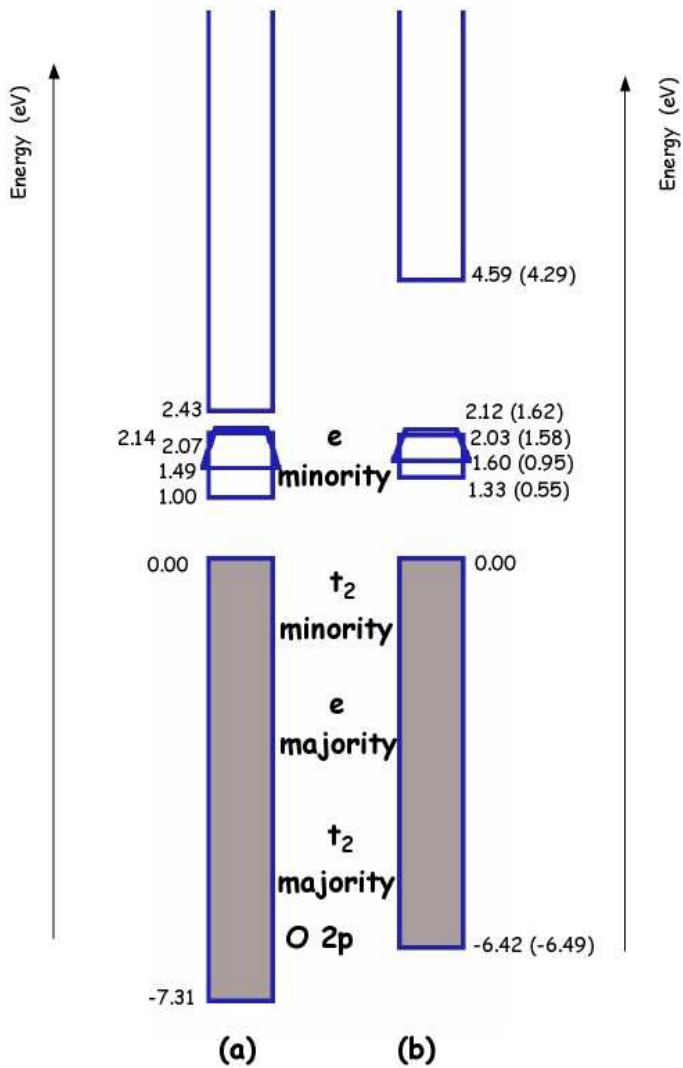}
\caption{\label{results} (a) NiO energy bands calculated with GGA.
(b) NiO energy bands calculated with GGA-1/2, and in parenthesis, 
calculated with LDA-1/2.}
\end{figure}

 In Fig.~\ref{results}(a) we show the valence and conduction bands
of the Kohn-Sham eigenvalues calculated with the PBE exchange-correlation
\cite{Perdew}. Contrary to the opinion of many authors, we do not see
anything wrong with those bands. The Kohn-Sham eigenvalues should
not compare well with the excitation spectrum, the band gaps being notoriously 
smaller. Unfortunately there seems to be no procedure to 
simultaneously calculate
the total energy, which is the object of Density Functional
Theory, and the excitation spectrum.
The spectrum calculation should be made following a very good GW \cite{Hedin}
that uses the self-energy operator instead of the exchange-correlation
potential. Though not correct, the Kohn-Sham energy spectrum usually gives
an indication of how the correct spectrum will be. As it stands in 
Fig.~\ref{results}(a) NiO is an insulator, the first conduction bands
are about 1~eV above the valence band and is made of minority Ni d 
wavefunctions. The identification of the many valence states as Ni~d 
majority and minority spins follows the simple argument: atomic Ni has 
2 electrons 4s and 8 electrons 3d. In NiO, the nickel looses the two
electrons 4s to the Oxygen, there completing its 2p shell. Due to the cubic
crystal field, the five 3d states of Nickel splits into $t_2$ with
degeneracy 3 and $e$ with degeneracy 2. Due to the trigonal distortion,
$t_2$ is further split into states with degeneracies 2 and 1. Of course
these identifications can only be seen at the zone center $\Gamma$.
Thus among the 8 remaining Ni valence electrons, 3 occupy $t_2$ with
majority spin, 2 occupy $e$ with majority spin, and 3 occupy $t_2$ with
a minority spin. The state $e$ with minority spin is empty and defines 
the first conduction band.
\par The first conduction bands are really acceptor bands. An electron
added to them would change the charge state of the Ni ion, according to
\[Ni^{++}+e\longrightarrow Ni^+\] 
These acceptor bands conduct very poorly and only by hopping from one Ni ion
to another with the same spin. They are responsible for
the dd excitation spectra of [\onlinecite{Muller}] and [\onlinecite{Huotari}],
at energies very different from the Sawatzky and Allen gap of 4.3~eV.
The band gap to these acceptor bands is only 1.00~eV, as calculated by
GGA \cite{Perdew}, or 0.51~eV as calculated with LDA \cite{LDA}. We must 
mention that one sees this gap even smaller in the literature due
to the use of plane wave bases. Anyway, though Fig.~\ref{results}(a)
gives the correct description of NiO, it misses the Sawatzky and Allen gap
altogether. The bands of NiO, described by Fig.~\ref{results}(a), and
the results of  Sawatzky and Allen remain incompatible.
\par Instead of a very precise GW calculation, Ferreira et al \cite{Ferreira}
developed a simple and very successful procedure, with no adjustable parameters,
to calculate the excitation energy spectrum. The procedure is inspired
in the old transition state technique for atoms, shown to be equivalent 
to the inclusion of the self-energy of the quasi-particle. Here it must
be recalled that the ``self-energy'' being calculated is not that of Hedin
\cite{Hedin}, but the classical self-energy of Electrostatics
\[\frac{1}{2}\int\int\frac{en(\vec{r})en(\vec{r}^{\prime})}
{|\vec{r}-\vec{r}^{\prime}|}d^3rd^3r^{\prime}\]
to which a minor exchange-correlation term is added. Thus the self-energy
operator of GW is expected to be equivalent to the 
exchange correlation potential
plus the self-energy potential defined in [\onlinecite{Ferreira}], and
plus an imaginary term that has eluded all calculations. As with the best
GW calculations, the Ferreira et al method, named LDA-1/2, produces
very good band gaps and effective masses, but at a very 
small computational price. 
Lately we have
been also using a GGA-1/2 method, where the exchange-correlation is
GGA, instead of LDA, and the -1/2 is to remind that we remove 1/2 
electron, as in the transition state technique. The self-energy of 
[\onlinecite{Ferreira}] has a very simple physical meaning: it is the work
necessary to join all the charge of the quasi-particle, dispersed in a
Bloch state, into a localized wavefunction. As such, this energy 
has to be subtracted from the Kohn-Sham eigenvalue. The self-energy
potential is calculated in the atom, and then trimmed so that it
does not extend to neighboring atoms. The trimming is made by means 
of a cutting function with a parameter ``CUT'' which is determined
variationally by making the band gap extreme. In the case of NiO we
used $CUT(Ni3d)=1.97a.u.$ and $CUT(O2p)=2.41a.u.$ to make the ``4.3 band gap''
maximum. The maxima are not sharp and neighboring values of $CUT$
could be used as well.
\par The energy spectrum calculated with the inclusion of self-energies
is represented in Fig.~\ref{results}(b). The results obtained with GGA does
differ from those from LDA. Aside from the relative upwards motion of
the ``4.3 band'', there is not much difference from the results
of a standard all-electron calculation (Fig.~\ref{results}(a)). It
is clear why the ``acceptor band'' at $\sim 1eV$ does not move
with respect to the valence bands: Both are Ni3d bands and have
the same self-energy.
On the other hand, these acceptor bands do not exist in the case of MnO.  
In this latter case, using our technique, we obtained a gap
of 4.18 ~eV (GGA) and 3.98~eV (LDA), matching the results of
other calculations and experiment, and no surprises were obtained
\cite{surprise}.
\par To conclude we want to say that standard LDA or GGA calculations
seem  reliable even for a strongly correlated material as NiO. But
it is important to remember that DF will never give the excitation spectrum.
An attempt to force the obtaining of such spectrum may generate errors
and misinterpretation of the calculated results, as
the impossible 4.3 eV Ni~d band.

\begin{thebibliography}{99}
\bibitem{surprise} With standard GGA, that is without our correction,
the band gap is 2.56~eV.
\end{thebibliography}
\end{document}